\documentclass[12pt]{elsarticle}
\usepackage{amssymb} 
	\usepackage{amsfonts}
	\usepackage[ansinew]{inputenc}
	\usepackage{epsfig}
 \usepackage{graphicx}
\begin{document}
\begin{frontmatter}
\title{A model for calcium-mediated coupling between the membrane activity and the clock gene expression in the neurons of the suprachiasmatic nucleus}
\author{J. M. Casado and M. Morillo}
%\email{casado@us.es}
\address{Area de F\'{\i}sica Te\'orica. Universidad de Sevilla\\
Apartado Correos 1085, 41080 Sevilla (Spain)}

\date{\today}

\begin{abstract}
Rhythms in electrical activity in the membrane of cells in the suprachiasmatic nucleus (SCN) are crucial for the function of the circadian timing system, which is characterized by the expression of the so-called clock genes. Intracellular Ca$^{2+}$ ions seem to connect, at least in part, the electrical activity of SCN neurons with the expression of clock genes. In this paper, we introduce a simple mathematical model describing the linking of membrane activity to the transcription of one gene by means of a feedback mechanism based on the dynamics of intracellular calcium ions.   
\end{abstract}

\end{frontmatter}
%\maketitle
\section{Introduction}
Circadian rhythms arise from the cooperative action of a number of endogenous biological oscillators generating daily patterns of many physiological and behavioral processes that persist even in the absence of the forcing provided by the external light-dark cycle. The master circadian clock in mammalians is located in the suprachiasmatic nucleus (SCN), a neuronal structure located in the anterior hypothalamus. The cells of the SCN behave as a set of cooperative autonomous oscillators with slightly distributed frequencies yielding a global circadian ($\sim$24-hour period) rhythm. Thus the basic oscillatory mechanism leading to the emergence of circadian rhythms has an intracellular origin that relies on the negative self-regulation of gene expression through transcriptional/translational feedback loops. In the last few years, a number of genes involved in such a regulatory mechanism has been identified \cite{COLWELL2011}. 

On the other hand, the neurons in the SCN show a characteristic firing pattern that changes dramatically with the circadian cycle. They are thought to encode the time of the day by adjusting their firing frequency to high rates during the day and lower ones at night.  Different studies carried out in recent times both in {\sl Drosophila} and in mammals suggest that the electrical activity of SCN cells provides the driving of the molecular clockwork. Also, keeping the SCN cells within an appropriate voltage range may be required for the generation of circadian rhythmicity of clock gene expression at the single cell level \cite{COLWELL2011}.  Thus, a fundamental question in circadian biology is how the electrical activity may regulate clock gene expression and, conversely, how this latter process may alter the electrical activity in SCN neurons.   

The evidence accumulated in recent years suggests that the effect of the electrical activity on clock gene expression by SCN cells is probably mediated, at least in part, by Ca$^{2+}$ ions. In fact, a close relationship between electrical activity and Ca$^{2+}$ levels has been observed. Resting levels of Ca$^{2+}$ in SCN neurons exhibit a circadian rhythm that has been detected by using a calcium sensitive dye. During the peak in firing at midday, SCN neurons show resting Ca$^{2+}$ levels of around 150 mM, but these levels drop to about 75 mM during times of inactivity. The action potential itself is an important source of Ca$^{2+}$ in the SCN, regulating Ca$^{2+}$ influx into the soma through the opening of voltage-sensitive calcium channels \cite{COLWELL2011}. This feature has been shown most clearly by a recent work in which a Ca$^{2+}$ levels and the firing of the SCN neurons were measured simultaneously \cite{IRWIN2007}. Data from this study show that driving the frequency of action potentials in the SCN neurons to 5-10 Hz (daytime levels) induces a rise in somatic Ca$^{2+}$ levels. This effect can be attenuated by the application of a L-type Ca$^{2+}$ channel blocker \cite{IRWIN2007}. 

In addition to the relative contribution of Ca$^{2+}$ influx to intracellular calcium levels, SCN neurons also have a rhythmically regulated reservoir of calcium that is not driven by membrane events. Both processes, inflow/outflow of calcium across the cell membrane and fixation/release of Ca$^{2+}$ from the calcium store, cooperate to generate intracellular Ca$^{2+}$ oscillations that seem to be crucial in driving a robust rhythm in gene expression. Indeed, rhythms in Ca$^{2+}$ levels, with its peak occurring during the day, seem to be a general feature of circadian systems \cite{COLWELL2011}. 

It turns out that the oscillatory behavior of the intracellular concentration of Ca$^{2+}$, and those of the variables of the genetic network as well, have a much slower time scale than that of the variables involved in the generation of action potentials by the cell membrane. Thus, from a theoretical point of view, the problem is to understand how fast variables can control the dynamics of slower ones and vice versa.  

A conductance based model of spiking in cells of SCN has recently been presented in \cite{SIM2007, BELLE2009}. In those works, the huge variety of ionic currents that contribute to the membrane excitability was modeled in terms of the sodium, potassium and leakage currents characteristic of the Hodgkin-Huxley formalism plus a specific calcium current. They describe the periodic alternation between silent and excited states of the neuronal membrane in terms of direct and inverse Hopf bifurcations induced by the externally imposed time dependence of the ionic conductances. Nonetheless, in those works, no attempt is made to link this driving to changes in the intracellular level of either Ca$^{2+}$ or of any other substance.  

The aim of this paper is to suggest a plausible mechanism linking the electrical activity in the cell membrane to the circadian rhythms of the genes expressions, taking into account the two very different time scales associated to the membrane voltage (fast) and the genetic (slow) processes. The basic ingredient is to admit that the intracellular level of Ca$^{2+}$ controls both the firing frequency and the transcription of a clock gene. The control of the firing frequency is achieved by the free Ca$^{2+}$ concentration that, in turn, is controlled by the Ca$^{2+}$ in the cell cytoplasm coming from reservoirs. The reservoir Ca$^{2+}$ concentration varies in a much longer time scale than the intracellular Ca$^{2+}$ one. At the same time, the intracellular Ca$^{2+}$ concentration evolves slowly compared to the other variables characterizing the membrane electrical activity. On the other hand, the control on the circadian variables is established by assuming that the 
activation of the clock gene expression is proportional to the free Ca$^{2+}$ concentration. 

\section{Ca$^{2+}$ controlled electrical activity}
We take the electrical activity taking place in the neuron membrane to be described by the two-dimensional system  
\begin{eqnarray}
\dot{x}&=&y-ax^{3}+bx^{2}+q-sz,\\
\dot{y}&=&c-dx^{2}-y.
\label{EQ001a}
\end{eqnarray}
The variable $x$ represents the (dimensionless) voltage across the cell membrane and $y$ is a recovering variable that describes the currents restoring the polarity of the membrane after the emission of an action potential, $q$ is a control parameter. In the first of these equations, $z$ stands for the free calcium concentration inside the cell, whose dynamics is supposed to be governed by the reactions
\begin{equation}
Z\stackrel{\epsilon k_{2}}{\longrightarrow} {\o},\qquad {\o}\stackrel{\epsilon k(x)}{\longrightarrow} Z\qquad Z_{R}\stackrel{\epsilon k_{3}}{\longrightarrow} Z.
\end{equation}
The first equation describes the outflow of Ca$^{2+}$ from the cell whereas the second one expresses the inflow of these same ions while the voltage-sensitive calcium channels are open during the membrane activation. Thus, the rate of this process is made dependent of the voltage. The last reaction describes the release of calcium ions from a intracellular reservoir described by the variable $z_R$. The changing in Ca$^{2+}$ levels induced by the reservoir seems to be a critical part of the output pathway by which intracellular processes drive rhythms in neural activity \cite{COLWELL2011}. The kinetic of calcium ions are completed by the reactions
\begin{equation}
Z\stackrel{\mu k_{4}}{\longrightarrow} Z_{R}\,\qquad Z_{R}\stackrel{\mu k_{5}}{\longrightarrow} {\o}
\end{equation}
where the first reaction describes the subtraction by the reservoir of free calcium ions and the second one corresponds to the direct loss of calcium ions by the reservoir. This set of reactions allows us to write the system of differential equations for the variables $z$ and $z_{R}$ 

\begin{eqnarray}
\dot{z}&=&\epsilon (k(x)-k_{2}z+k_{3}z_{R}),\\
\dot{z}_{R}&=&\mu(k_{4}z-k_{5}z_{R}).
\label{EQ001b}
\end{eqnarray}

If we admit that the temporal scale for the evolution of the reservoir is much slower that the corresponding to the free calcium ($\mu\ll\epsilon$), we have
\begin{equation}
\dot{z}_{R}=0,\quad\Rightarrow\quad z_{R}=\Gamma,
\end{equation}
$\Gamma$ being constant on the $\epsilon$ time scale. Thus, introducing this constant in Eq. \ref{EQ001b} and assuming that $k(x)=k_{1}x$, we can write 
 \begin{eqnarray}
\dot{x}&=&y-ax^{3}+bx^{2}+q-sz,\\
\dot{y}&=&c-dx^{2}-y,\\
\dot{z}&=&\epsilon(k_{1}x-k_{2}z+g),
\label{EQ002}
\end{eqnarray}
with $ g = k_3 \Gamma$.
This three-dimensional dynamical system is formally identical to the Hindmarsh-Rose (HR) neuronal model \cite{HINDMARSH1982}. Here, however, the {\sl slow adaptation variable} $z$ introduced by these authors is reinterpreted as the intracellular free Ca$^{2+}$ concentration, whose level controls the much faster spiking variables $x$ and $y$. The HR model shares essential qualitative features with several conductance-based models of excitable cells giving rise to square-wave bursting \cite{CHAY1983,BUTERA1999,GOLOMB2006}. All those models involve the interplay between a stable branch of stationary points and a limit cycle through a saddle middle branch in the equilibrium curve. By inducing the periodical transitions between these two coexisting attractors by means of a slow control variable, this dynamical mechanism allows the description of periodical silent and active phases in the neuron behavior. This same dynamical mechanism has recently been invoiced to interpret some experimental findings related to crustacean patterns generators \cite{MARIN2014}.
     
\begin{figure}[h]
\centerline{\epsfig{figure=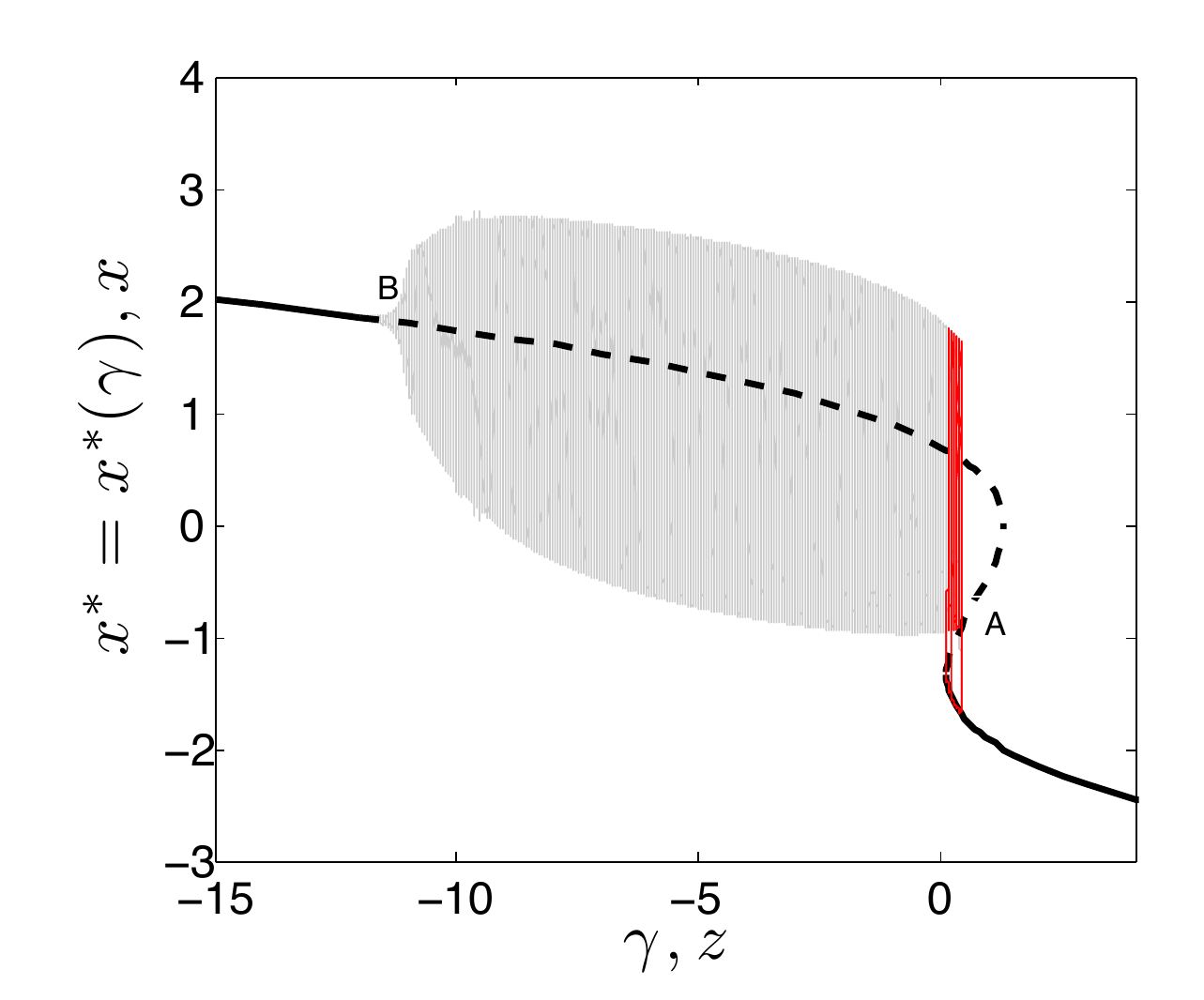,height=6cm}}
%\centering\includegraphics[width=0.4\linewidth]{Fig1.pdf}
\caption{\small $Z-$shaped steady state curve $x^{\ast}=x^{\ast}(\gamma)$ for the fast subsystem (Eq.\ref{EQ003}). The stable branch is depicted by a continuous line whereas the dashed lines represent unstable branches of equilibria. The limit cycle appearing at $\gamma=-11.293$  and ending at the homoclinic bifurcation at $A$ appears in gray. The projection of the global attractor of the full three-variable system (in red) has been superimposed to this diagram. Parameter values are $a=1.0, b=3.0, c=1.0, d=5.0, q=0.3, k_{1}=1.0, k_{2}=0.1$ and $g=1.33$.}
\label{FIG1}
\end{figure}

The HR system with $\epsilon\ll 1$ has two very different time scales, one of them associated with a {\sl fast subsystem} ($x$ and $y$ variables) and the other associated with the dynamics of a {\sl slow subsystem} ($z$ variable). In fact, $\epsilon\ll 1$ implies that in the time scale in which both $x$ and $y$ vary, $z$ can be considered as a constant. Thus, calling $sz=\gamma$ we are led to analyze the dynamics of the two-dimensional system
 \begin{eqnarray}
\dot{x}&=&y-ax^{3}+bx^{2}+q-\gamma,\nonumber\\
\dot{y}&=&c-dx^{2}-y,
\label{EQ003}
\end{eqnarray}
in terms of the values of the parameter $\gamma$. The equilibrium values $x^{\ast}$ of this two-dimensional system obey a cubic equation resulting from the crossing of both nullclines associated with Eq.(\ref{EQ003}), 
\begin{equation}
ax^{3}+(d-b)x^{2}-q-c+\gamma=0.
\label{EQ004}
\end{equation}
Several regimes can be found as the parameter $\gamma$ is varied. As seen in Fig.\ref{FIG1} for small negative values of $\gamma$ the only attractor is an equilibrium point (a stable focus) with non-zero $x^{\ast}$-values. As $\gamma$ increases,  the system undergoes a Hopf bifurcation rendering unstable the branch of equilibrium points giving rise to a stable limit cycle corresponding to voltage spikes (point $B$ in Fig.\ref{FIG1}). For still larger $\gamma$ values, the cubic equation has two unstable values coexisting with a stable one. Furthermore, the limit cycle disappears through a homoclinic bifurcation when it collides with the unstable branch of $x^{\ast}$ (point $A$ in the same figure). Notice the rather small interval of $\gamma$ values for which the stable limit cycle coexists with the stable equilibrium point appearing through a saddle-node bifurcation. For still larger values of $\gamma$, just a stable stationary branch remains.

\begin{figure}[h]
\centerline{\epsfig{figure=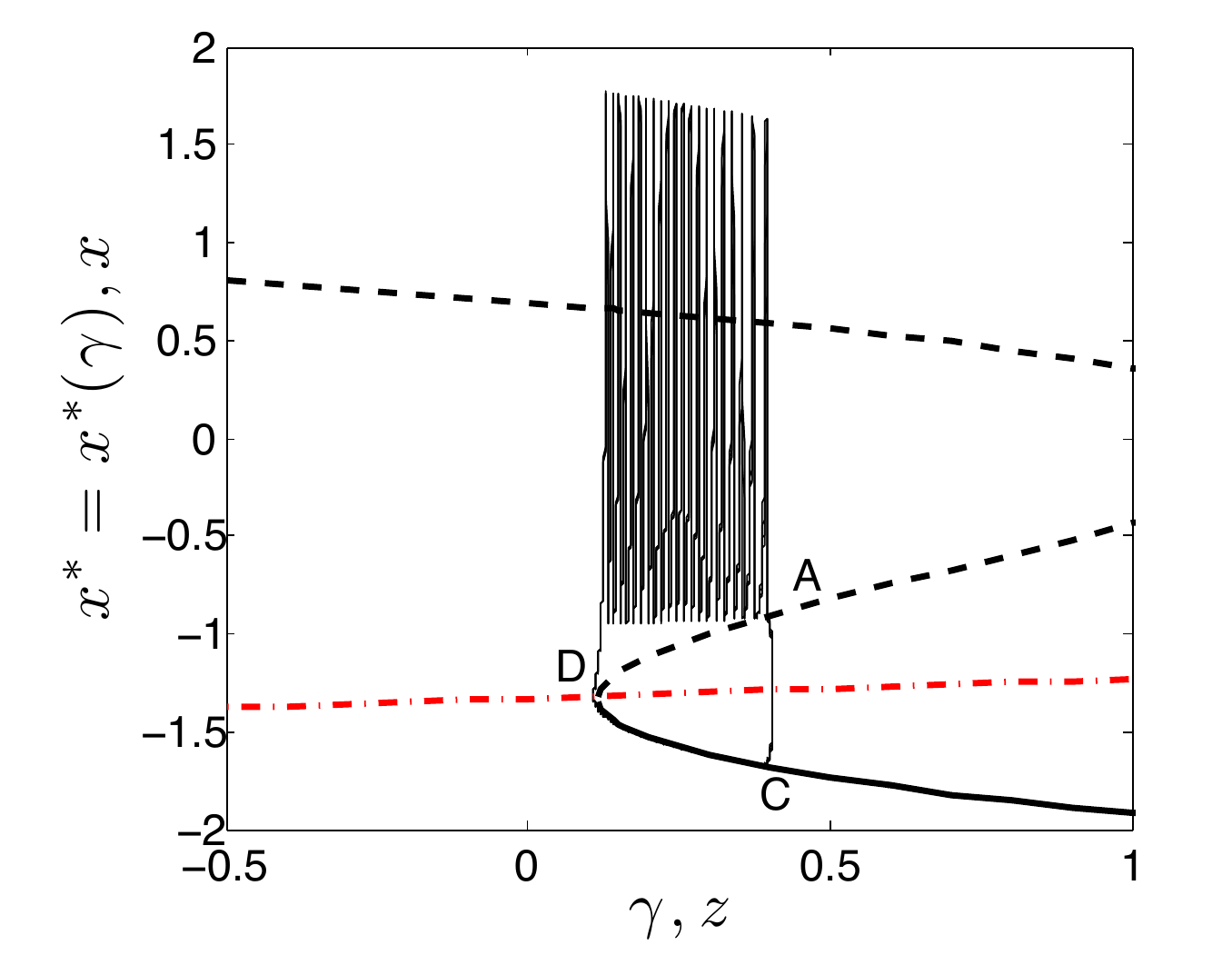,height=6cm}}
\caption{\small Enlarged version of coexistence region depicted in Fig. \ref{FIG1}. The homoclinic bifurcation occurs at $A$ and the fold bifurcation at $D$. The red dot-dashed line corresponds to the $z-$nullcline. The solid black thin line corresponds to the bursting oscillations depicted in red in Fig. \ref{FIG1}.}
\label{FIG2}
\end{figure}

Following a technique pioneered by J. Rinzel some decades ago \cite{RINZEL1985} a rather useful picture of the HR model dynamics as given by Eq. (\ref{EQ002}) can be obtained by projecting its orbits onto the $xz$-plane, and superimposing it with the bifurcation diagram for the fast subsystem described by Eq. (\ref{EQ003}).
In Fig.\ref{FIG2}, an enlarged version of the coexistence region is presented including the nullcline of the slow subsystem in (\ref{EQ002}) $x=(k_{2}z-g)/k_{1}$ (the red dot-dashed line). The solid black thin line in this figure corresponds to the bursting oscillations depicted in red in Fig. \ref{FIG1}. This curve has been obtained from the $x(t)$ and $z(t)$ obtained by numerically integrating the full system of differential equations (\ref{EQ002}). It corresponds to the projection on the $x-z$ plane of the global three dimensional attractor.
 

As one can observe in Fig. \ref{FIG2}, when $z$ slowly moves to the left (notice that $\dot{z}<0$ under the $z-$nullcline), the $x$ variable evolves along the branch of stable equilibria ($C\rightarrow D$) until it comes to an end at a saddle-node bifurcation (point $D$). At this point it is forced to jump to the limit cycle by crossing the $z-$nullcline and afterwards it is forced to move to the right (now $\dot{z}>0$) while performing fast oscillations. This oscillatory behavior ends when the limit cycle disappears through the homoclinic bifurcation at $A$. At this point,  the $x$ variable must jump to the stable branch of equilibria and the whole process repeats itself. 

The temporal evolution of $x(t)$ and $z(t)$ has been depicted in Fig.\ref{FIG3}. As we can see, the free running $x(t)$ variable alternates periodic episodes of fast spiking with silent epochs of much slower evolution. On the other hand, $z(t)$ evolves in the slower time scale. In this case, the temporal scale of the transitions between the silent and the spiking states of the neuron coincides with that of the $z$ variable. Each period of the $z$ oscillations corresponds to a cycle characterized by a rise of the calcium level associated with the burst of spikes and followed by a silent phase in which the concentration of Ca$^{2+}$ slowly decays.  

Note that our model describes the membrane activity without explicitly incorporating the ionic currents.  Nonetheless, our equations show a bifurcation structure close to that of more complicated conductance models \cite{BELLE2009}. 

\section{Coupling electrical and molecular activities}
On the genetic side of the model, we have considered the transcription of only one clock gene. Thus the variables of interest are the intracellular concentration of clock gene mRNA, the corresponding level of a clock protein and that of a transcriptional inhibitor whose action closes the self-regulatory feedback loop. All these variables obey the differential equations of the well-known Goodwin model \cite{WOLLER2013}. 

To link the membrane electrical activity to the clock gene expression, we assume that the level of free calcium acts as an activator of the transcription of the mRNA. Then, our complete model is embodied in the six-variable system of differential equations
\begin{eqnarray}
\dot{x}&=&y-ax^{3}+bx^{2}-sz+q+pY,\\
\dot{y}&=&c-dx^{2}-y,\\
\dot{z}&=&\epsilon(k_{1}x-k_{2}z+g),\\
\dot{X}&=&\epsilon\Big(\frac{\alpha z}{1+Z^{h}}-kX\Big),\\
\dot{Y}&=&\epsilon(k_{f}X-kY),\\
\dot{Z}&=&\epsilon(k_{f}Y-kZ)
\label{EQ0066}
\end{eqnarray} 
where $X, Y$ and $Z$ describe the clock mRNA, the clock protein and the inhibitor of the transcription, respectively. Throughout this work we will take $a=1, b=3, c=1, d=5, s=1, h=10, k=2, k_{f}=2$ and, $\alpha=8$. Note that the above equations contemplate a direct feedback mechanism by which the molecular clock is able to drive the electrical activity of the cell membrane through the term $pY$ appearing in the voltage equation. At the same time, the variable $z$ representing the free Ca$^{2+}$ concentration inside the cell, influences the clock gene concentration $X$ through a Hill-like term.  For more elaborated circadian models involving several clock genes \cite{BW,LG} the coupling between the membrane and genetic activities could be formulated along the lines  of our methodology by including terms describing the Ca$^{2+}$ activation of the different gene expressions.

\begin{figure}[h]
\centerline{\epsfig{figure=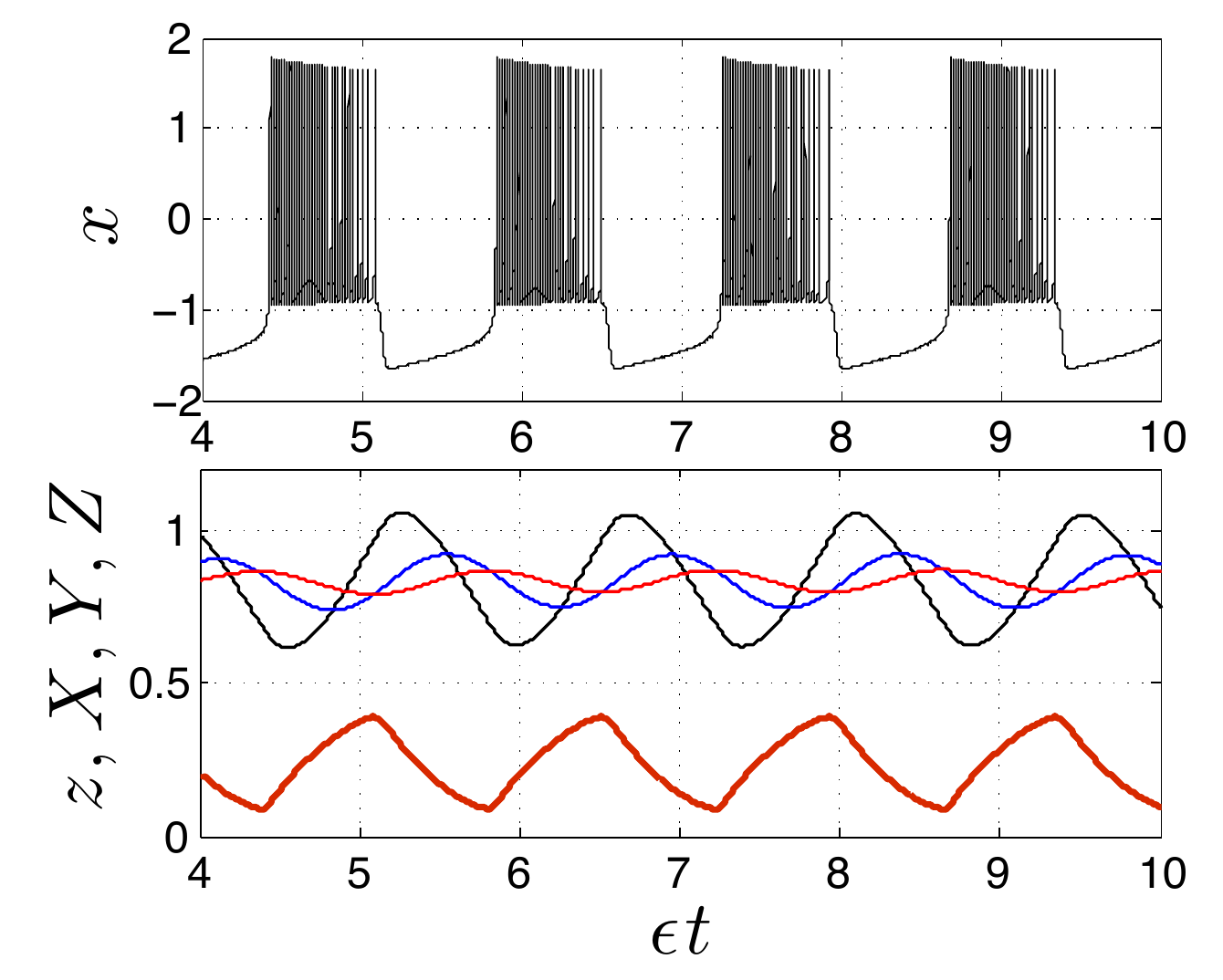,height=6cm}}
\caption{\small Square-wave bursting of the model with $p=0$ as a function of (dimensionless) time $\epsilon t$ (upper panel).The evolution of $X(t)$ (black), $Y(t)$ (blue) and $Z(t)$ (red) as well as the temporal behavior of $z(t)$ (red thick line) have been depicted in the lower panel. Parameter values are $\epsilon=0.001$, $p=0, q=0.3$, $k_{1}=1.0$, $k_{2}=0.8$ and $g=1.23$.}
\label{FIG3}
\end{figure}

We first study the case with no feedback from the genetic variables on the membrane ones ($p=0$). In Fig. \ref{FIG3} we have depicted the temporal behavior of the voltage variable and that of the intracellular concentration of free calcium as well as the evolution of all the variables of the genetic subsystem. The parameter $\epsilon$ sets the spiking frequency of the system so that the smaller it is, the higher the frequency of spiking achieved during the bursting phase. The value of $q$ is critical to the character and persistence of these bursts. 
In order to save computation time we have set the values of $\epsilon $ small but with no intention to fit quantitatively the experimental values for the neurons of the SCN. One could interpret the behavior observed in Fig. \ref{FIG3} by noting that the start of a burst of spikes activates the inflow of Ca$^{2+}$, thus increasing its intracellular level until it forces the membrane back to its resting electrical state. On the other hand, the slow oscillation in the intracellular calcium concentration forces a slightly delayed oscillation in the level of mRNA (black line in the lower panel), leading to an additional delay in the evolution of the protein level (blue line) as well as in the inhibitor (red line). Note that there is no modulation of the bursts amplitude due to the fact that we have chosen $p=0$. 

\begin{figure}[h]
\centerline{\epsfig{figure=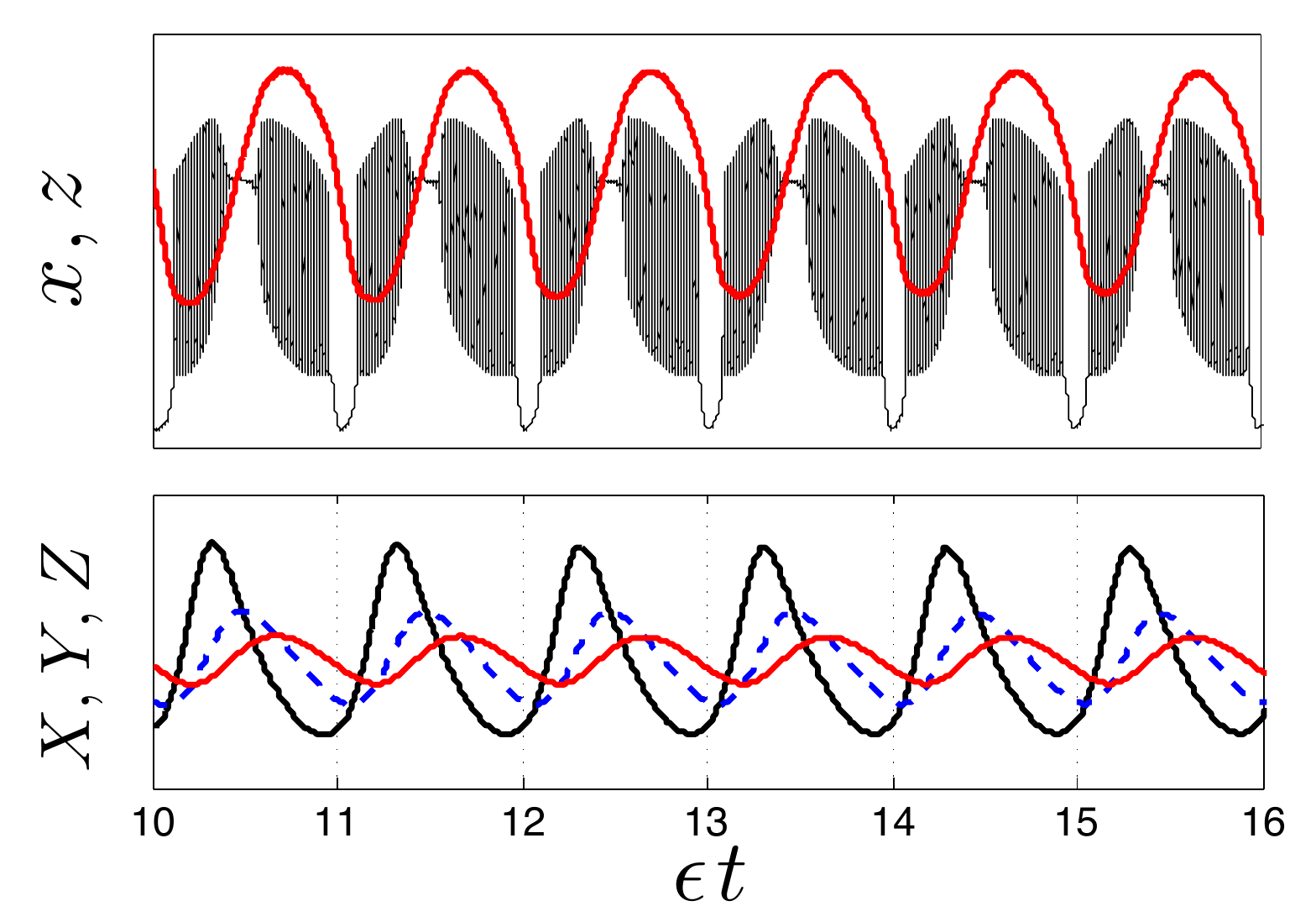,height=6cm}}
\caption{\small Autonomous behavior of $x(t)$  and $z(t)$ for the model with $p\ne 0$ (upper panel). The evolution of $X(t)$ (black), $Y(t)$ (blue) and $Z(t)$ (red) have been depicted in the lower panel. Parameter values are $k_{1}=4$, $k_{2}=0.6, q=0.5$ and $p=12.5$.}
\label{FIG4}
\end{figure}

\begin{figure}[ht]
\centerline{\epsfig{figure=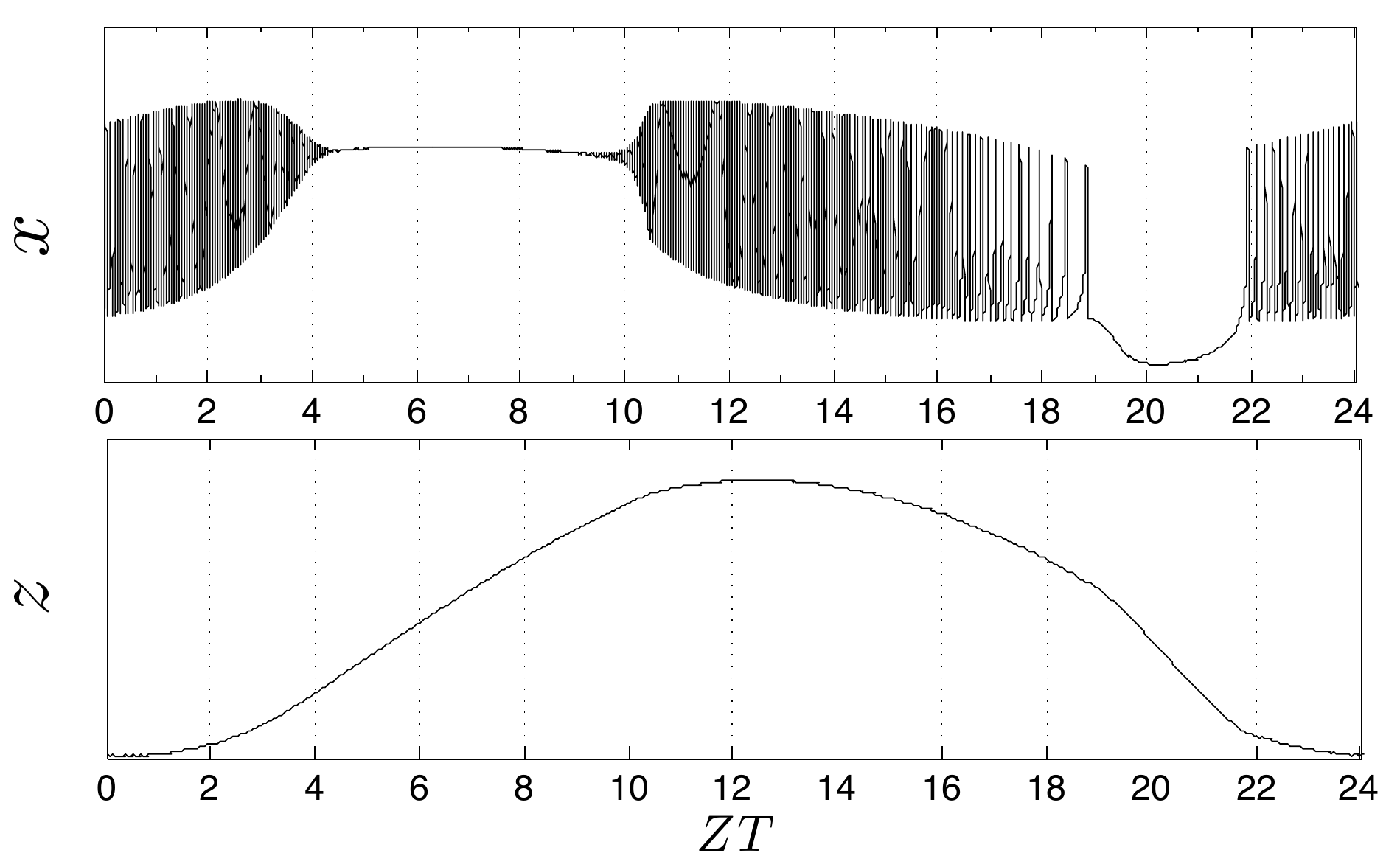,height=6cm}}
\caption{\small Autonomous behavior of the voltage variable (upper panel) and that of the concentration of intracellular calcium (lower panel) during a whole day. The silent phase characterized by a strongly depolarized membrane appears through an inverse Hopf bifurcation and ends at a direct bifurcation of the same kind. Parameter's values are $k_{1}=4$, $k_{2}=0.6$, $q=0.5$, and $p=13.5$. The time origin has been shifted to coincide with that used in \cite{BELLE2009}.}
\label{FIG5}
\end{figure}

Now we turn to the case with $p\ne 0$. As we can see in Fig.\ref{FIG4}, the existence of the $pY$ feedback term in the first equation of the model generates a relatively complex temporal modulation of the voltage variable along the day. Again, the genetic variables oscillate at the slower time scale of the voltage variable and with a certain dephasing among them. As we can see in Fig.\ref{FIG4}, the level of the clock gene mRNA peaks at the start of the silent phase of the neuron that takes place during the daylight hours (see Fig.\ref{FIG5}) while it decreases during the night.

In Fig.\ref{FIG5} we present the autonomous behavior of the voltage variable in terms of the Zeitgeber Time (ZT) during a whole day. The time origin for the data plotted in this figure has been shifted to coincide with that of the data reported by Belle and coworkers for the forcing of {\sl per1}-containing SNC neurons with a sequence of equal periods of light and darkness \cite{BELLE2009}. The results presented here are qualitatively close to those experimental findings although some differences in timing remains. From ZT0 to ZT2 approximately, the neuron is firing, whereas from ZT4 to ZT10, it has been deeply depolarized and has become silent. The value chosen for $p$ in the equations of the model affects the period of depolarization. For $p=13.5$, the spiking behavior starts again at ZT10 and lasts until ZT18, when the neuron becomes hyperpolarized and its activity ceases. This silent period increases as the parameter $k_{2}$ decreases. Finally, at Z22 the neuron starts firing again. The intracellular concentration of calcium ions, on the other hand, rises during the day, peaks in the evening and early night and decreases to a minimum along the late night hours. Note that the spiking frequency changes with the ZT value.

\section{Concluding remarks}
Experimental evidence indicates that the link between electrical activity and clock gene expression in SCN cells is provided, at least in part, by the intracellular dynamics of Ca$^{2+}$ ions associated with the existence of a reservoir of these ions inside the cell. The nature and localization of this 
intracellular reservoir is unknown at present although it seems that this role is played by the endoplasmic and sarcoplasmic reticulae \cite{NOBLE01}. 

In this work we have presented a mathematical model aiming to explore a simple dynamical mechanism for the driving of clock genes by the firing activity of neurons.  Some work carried out in the last few years suggests the existence of a control of the firing frequency by clock gene expression, but very little is known about the mechanisms by which genetic oscillations are able to drive rhythms in neural activity \cite{COLWELL2011}. So, in this work we have explored this issue by assuming that the protein produced by the transcription of the clock gene modulates the neuronal firing. Similar results would have been obtained by using the mRNA variable or the inhibitor to control the neuronal firing. The role played by a calcium reservoir has been taken into account by considering the kinetics of calcium fixation and its release by the reservoir. It seems possible to extend the main ideas that we have elaborated here to link the genetic and membrane variables of the SNC using a conductance based model for the membrane activity as well as a dynamics of more than one clock gene. Work along these lines is in progress.

\end{document}